\documentclass[12pt]{article}
\def\bea{\begin{eqnarray}}
\def\eea{\end{eqnarray}}

\def\ba{\beq\new\begin{array}{c}}
\def\ea{\end{array}\eeq}

\newcommand{\beq}{\begin{equation}}
\newcommand{\eeq}{\end{equation}}
\newcommand{\beqa}{\begin{eqnarray}}
\newcommand{\eeqa}{\end{eqnarray}}

\newcommand{\Tr}{{\rm Tr}\,}

\newcommand{\ntwo}{\mbox{${\cal N}\!\!=\!2\;$}}
\newcommand{\none}{\mbox{${\cal N}\!\!=\!1\;$}}

\begin{document}
%\begin{titlepage}
\renewcommand{\thefootnote}{\fnsymbol{footnote}}

\begin{flushright}
ITEP-TH-53/00\\
hep-th/0009206\\

\end{flushright}

\vfil

\begin{center}
\baselineskip20pt
{\bf \Large Condensates near the Argyres-Douglas point in\\ SU(2)  gauge
theory with  broken \mathversion{bold}\ntwo  supersymmetry }
\end{center}

\vfil

\begin{center}
{\large A. Gorsky$^a$}

\vspace{0.3cm}

$^a${\it Institute of Experimental and Theoretical Physics, Moscow
117259,}\\[0.2cm]

\vfil

{\large\bf Abstract} \vspace*{.25cm}
\end{center}

The behaviour of the chiral condensates in the SU(2) gauge theory
with broken \ntwo super\-symmetry is reviewed.  The
calculation of monopole, dyon, and charge condensates is described.
It is  shown that the monopole and charge condensates  vanish at  the
Argyres-Douglas point where the monopole and charge vacua collide.
This phenomenon is interpreted as a deconfinement of electric and magnetic
charges at  the Argyres-Douglas point.

\vfil
\begin{flushleft}
 Talk given at "Quarks-2000", Pushkino, May 2000
\end{flushleft}
%\end{titlepage}

\section{Introduction}
This talk is based on the paper \cite{gvz} where the behaviour
of the supersymmetric gauge theories near the Argyres-Douglas point
was considered. The main question discussed concerned the nature
of the hypothetical phase transition occured at the Argyres-Douglas
point. It is widely believed that the theory near this point
flows into the superconformal point in the infrared however
the physics of this critical point was unclear. Since the results
presented below are based on the exact statements concerning
\none and \ntwo supersymmetric theories the
identification of the phase transition at the Argyres-Douglas point
as a deconfainment phase transition is rigorous.

The derivation of exact results in \none supersymmetric gauge theories
based  on low energy effective superpotentials and holomorphy was pioneered
in~\cite{ads,SV} and then strongly developed, mostly by Seiberg,
see~\cite{IS} for review.
An extra input was provided by the Seiberg-Witten solution
 of \ntwo supersymmetric gauge theories with and without
matter~\cite{sw}.
It was also clarified that Seiberg-Witten solution
amounts the set of vacua in
the corresponding \none theory~\cite{sw,is,giveon,kitao,kt}.
Different vacua are distinguished by values of chiral condensates, such
as  gluino condensate $\langle\Tr\lambda \lambda\rangle$ and the
condensate  of the fundamental matter $\langle\tilde{Q} Q\rangle$.
Recently some points
concerning  the formation
of the condensate and the identification of the relevant field
configurations
were clarified in \cite{hol,davies,rv,konishi}

We compare then the condensate of the adjoint matter with the
discriminant  locus
defined by Seiberg-Witten solution in \ntwo theory and find a complete
matching.
Our results for matter and gaugino condensates are consistent with those
obtained
by `integrating in' method~\cite{ils,Intrilligator,giveon} and can be
viewed  as an
independent confirmation of the method.
What is specific for our approach is that
we start from  weak coupling regime where notion of effective
Lagrangian is  well
defined and  then use holomorphy to extend results for chiral condensates into
strong coupling.

Then we determine monopole, dyon and charge condensates following to
the Seiberg-Witten approach, i.e. considering
effective superpotentials near singularities on the Coulomb branch in \ntwo
theory. Again, holomorphicity allows us to extend results to the
domain of strong \ntwo breaking.

Our next step is study of chiral condensates  in the Argyres-Douglas (AD)
points. These points were originally introduced in moduli/parameter
space of \ntwo theories as
points  where two singularities on the Coulomb branch
collide~\cite{ad,apsw,hori}. It is believed that the theory at
the AD point flows in infrared to a nontrivial
superconformal theory. The notion of AD point continue to
make sense even when the  \ntwo theory  is broken to \none by nonzero
$\mu$, in the \none theory it is the point in parameter space where
 two vacua collide.

Particularly, we consider collision of monopole and charge vacua at
certain value of the mass of the fundamental flavor. Our key result is
that  both monopole and charge condensates  vanish at the
 AD  point. We interpret this  as deconfinement of both
electric and magnetic charges at the  AD
point.

Let us remind that the condensation of monopoles ensures confinement
of quarks in the monopole vacuum~\cite{sw}, while the condensation of
charges provides confinement of monopoles in the charge vacuum.
As it was shown by 't~Hooft~\cite{H} it is impossible for these two
phenomena  to coexist. This leads to a paradoxical situation in the AD
 point where the monopole and charge vacua collide.
Our result resolves this paradox.

This paradox is a part of more general problem: whether there is a uniquely
defined theory in the AD point. Indeed, when two vacua collide the
Witten index of the emerging theory is 2, i.e. there are two bosonic
vacuum states. The question is if there is any physical quantity which
could serve as an order parameter differentiating these two vacua.
The continuity of chiral condensates in the AD point we found
shows that these condensates are not playing this role. The same continuity
leads also to vanishing of tension of domain walls interpolating
between colliding vacua when we approach the AD point.

\section{Matter and gaugino  condensates }\label{sec:matgaug}
%\subsection{Effective superpotential and condensates}
\label{sec:sup}

Let us consider \none theory with SU(2) gauge group where the matter
sector consists of the adjoint field $\Phi^\alpha_\beta=\Phi^a
(\tau^a/2)^\alpha_\beta$ ($\alpha,\beta=1,2;~a=1,2,3$) and two fundamental
fields
$Q^\alpha_f$ $(f=1,2)$ describing one flavor.
The most general renormalizable superpotential for this theory has the form,
\begin{equation}
{\cal{W}}= \mu \,{\rm Tr} \,\Phi^2 + \frac{m}{2} \,Q_{ f}^\alpha Q^{f}_\alpha
+\frac{1}{\sqrt{2}}\,h^{fg} \,Q_{\alpha f}\Phi^\alpha_\beta Q^\beta_g\;.
\label{superp}
\end{equation}
Here parameters $\mu$ and $m$ are related to masses of the adjoint and
fundamental fields,
$m_\Phi=\mu/Z_\Phi$, $m_Q=m/Z_Q$, by corresponding $Z$ factors in kinetic
terms. Having  in mind normalization to the \ntwo case we choose for bare
parameters
$Z_\Phi^0=1/g_0^2$, $Z_Q^0=1$. The matrix of Yukawa couplings
 $h^{fg}$ is the symmetric,  summation over color indices $\alpha,\beta=1,2$
is explicit. Unbroken \ntwo SUSY appears when $\mu=0$ and $\det h=-1$.

To get an effective theory similar to SQCD we integrate out the adjoint field
$\Phi$ implying  that $m_\Phi\gg m_Q$. In classical approximation this
integration
reduces to to the substitution
\begin{equation}
\Phi^\alpha_\beta= -\frac{1}{2\sqrt{2}\,\mu}
\,h^{fg}\left( Q_{\beta f}Q^\alpha_g
-\frac 1 2 \delta^\alpha_\beta  Q_{\gamma f}Q^\gamma_g\right)\,,
\label{substitute}
\end{equation}
which follows from $\partial {\cal{W}}/\partial \Phi =0$.
It is well known from the
study of SQCD that perturbative loops do not contribute and nonperturbative
effects
are exhausted  by the  Affleck-Dine-Seiberg (ADS)
superpotential generated by one instanton~\cite{ads}.
The
effective superpotential then is
\begin{equation}
{\cal{W}}_{\rm eff}= m\, V - \frac{(-\det h)}{4\mu}\, V^2 +\frac{\mu^2
\Lambda_1^3}{4\,V}
\label{sup1}
\end{equation}
where the gauge and subflavor invariant chiral field $V$ is defined as
\begin{equation}
V=\frac 1 2 \,Q_{ f}^\alpha Q^{f}_\alpha \;.
\end{equation}
The third nonperturbative term in
Eq.~(\ref{sup1}) is the  ADS
superpotential.  The coefficient $\mu^2 \Lambda_1^3/4$ in the ADS superpotential is an
equivalent of $\Lambda^5_{\rm SQCD}$ in SQCD.  The factor $\mu^2$ in the
coefficient reflects four zero modes of the adjoint field, see e.g.
Ref.~\cite{y,rv} for
details.

When $\det h$ is nonvanishing we have three vacua, marked by vevs of
the lowest component of $V$,
\begin{equation}
v=\langle \,V \,\rangle\,.
\end{equation}
These vevs are roots of
 the algebraic equation ${\rm d}{\cal W}_{\rm eff}/{\rm d}v=0$ which looks as
\begin{equation}
m-\frac{(-\det h)}{2}  \, \frac{v}{\mu}- \frac{\Lambda_1^3}{4} \left(
\frac{\mu}{v}\right)^2=0\;.
\label{vaceq}
\end{equation}
This equation shows, in particular, that although the second term in the
superpotential (\ref{sup1}) looks as suppressed at large $\mu$
it is of the same
order as the ADS term. From Eq.~(\ref{vaceq}) it is also clear that the
dependence
on
$\mu$ is  given by scaling $v\propto\mu$.

To see dependence on other
parameters let us substitute $v$ by the dimensionless variable $\kappa$ as
\begin{equation}
v=\mu\,\sqrt{\frac{\Lambda_1^3}{4m}}\,\kappa\;.
\label{vk}
\end{equation}
Then Eq.~(\ref{vaceq}) in terms of $\kappa$
\begin{equation}
1-\sigma\, \kappa -\frac{1}{\kappa^2}=0
\label{kappa}
\end{equation}
is governed by the dimensionless parameter $\sigma$,
\begin{equation}
\sigma= \frac{(-\det h)}{4} \,\left(\frac{\Lambda_1}{m}\right)^{3/2}.
\label{sigma}
\end{equation}
.

To verify this interesting mapping we need to find out vevs for
\begin{equation}
u=\langle U\rangle=\langle{\rm Tr} \,\Phi^2\rangle\;.
\end{equation}
This can be done using set of Konishi anomalies.
Generic equation for arbitrary matter field $Q$ looks as follows (we are using
notations of the review~\cite{sv}):
\begin{equation}
\frac{1}{4}\,\bar{D}^{2}J_Q= Q\,\frac{\partial \,{\cal{W}}}{\partial
Q}+T(R)\,\frac{{\rm Tr}\,W^2}{8\pi^2}\;,
\end{equation}
where $T(R)$ is the Casimir in the matter representation. The left
hand side is the total derivative in superspace so its average over
supersymmetric
vacuum vanishes. In our case it results in two relations for condensates,
\begin{eqnarray}
\left\langle \frac{m}{2}\,Q_{ f}^\alpha Q^{f}_\alpha +\frac{1}{\sqrt{2}}\,h^{fg}
\,Q_{\alpha f}\Phi^\alpha_\beta Q^\beta_g+\frac12 \,\frac{{\rm Tr}\,W^2}{8\pi^2}
\right\rangle=0 \nonumber\\
\left\langle 2\,\mu {\rm Tr} \,\Phi^2 + \frac{1}{\sqrt{2}}\,h^{fg} \,Q_{\alpha
f}\Phi^\alpha_\beta Q^\beta_g +2\,\frac{{\rm Tr}\,W^2}{8\pi^2}
\right\rangle=0
\label{konishirel}
\end{eqnarray}
From the first relation after substitution (\ref{substitute}) and comparison
with
Eq.\ (\ref{vaceq}) we find the expression for gluino condensate $s$
\begin{equation}
s=\frac{\langle {\rm Tr}\,\lambda^2\rangle}{16\pi^2}=-\frac{\langle {\rm
Tr}\,W^2\rangle}{16\pi^2}=\frac{\mu^2 \Lambda_1^3}{4\,v}\;.
\end{equation}
This is consistent with the general expression $[T_G -\sum T(R)]\langle {\rm
Tr}\lambda^2\rangle/16\pi^2$ for the nonperturbative ADS piece of the
superpotential (\ref{sup1}) \cite{NSVZ}. Combining then two relations
(\ref{konishirel}) we express  the condensate value of $u$ via $v$,
\begin{equation}
u=\frac{1}{2\mu}\left(m\,v +3\,s\right)=
\frac{1}{2\mu}\left(m\,v +\frac 3 4\,\frac{\mu^2 \Lambda_1^3}{v}
\right)=
\frac{\sqrt{m\Lambda_1^3}}{4}\left(\kappa+\frac{3}{\kappa}\right)\;.
\label{ukappa}
\end{equation}
Now we see that at the limit of large $m$ two vacua $\kappa=\pm 1$ are in
perfect
correspondence with $u=\pm \,\Lambda_0^2$ for the monopole and dyon vacua of
SYM. Indeed,
$\Lambda_0^4=m\Lambda_1^3$ is a correct relation between scale parameters of
the theories.

For the third vacuum at large $m$ the value $u=m^2/(- \det h)$
corresponds on the Coulomb branch to the so called charge vacuum, where
some fundamental fields become massless.  Moreover,  the correspondence
with \ntwo results can be demonstrated for three vacua at any value of $m$. To
this end  we use the relation (\ref{ukappa}) and Eq.~(\ref{kappa}) to derive the
following equation for
$u$,
\begin{equation}
(-\det h)\,u^3-m^2\,u^2-\frac 9 8 \,(-\det
h)\,m\Lambda_1^3\,u+m^3\Lambda_1^3 +\frac{27}{2^8}\,(-\det
h)^2\Lambda_1^3=0\;.
\label{uone}
\end{equation}
Three roots of this equation are vevs of $\Tr \,\Phi^2$ in the corresponding
vacua.

How does it look from \ntwo side?  The Riemann surface
governing the Seiberg-Witten solution is given by the curve \cite{sw}
\begin{equation}
y^2=x^3-u\,x^2+ \frac{1}{4}\Lambda_{1}^3 m\,x-\frac{1}{64}\Lambda_{1}^{6}\;.
\label{swu}
\end{equation}
Singularities of the metric, i.e. the discriminant locus of the curve, is
defined by
two equations, $y^2=0$ and ${\rm d}y^2/{\rm d}x=0$,
\begin{equation}
x^3-u\,x^2+ \frac{1}{4}\Lambda_{1}^3 m\,x-\frac{1}{64}\Lambda_{1}^{6}=0\;,
\quad 3x^2 -2u\,x + \frac{1}{4}\Lambda_{1}^3 m=0\;,
\end{equation}
which lead to
\begin{equation}
u^3-m^2\,u^2-\frac 9 8 \,m\Lambda_1^3\,u+m^3\Lambda_1^3
+\frac{27}{2^8}\,\Lambda_1^3=0\;.
\label{utwo}
\end{equation}
We see that this is a particular case of the \none equation (\ref{uone})
at $\det h=-1$.

%\subsection{Argyres-Douglas points}

The point in the parameter manifold where two vacua coincide is the AD
 point~\cite{ad}.
In $SU(2)$ theory these points were studied in \cite{apsw}.
Mutually non-local states, say charges and monopoles becomes
massless at these points. On the Coulomb branch of \ntwo theory
these points correspond to non-trivial conformal field theory
\cite{apsw}. Here we study   the \none SUSY theory, where \ntwo is broken down
by the mass term for the adjoint matter as well as by the difference of the
Yukawa coupling from its \ntwo value.  But  collisions of two vacua
still occur in the theory.  In this subsection we  find the values of $m$
at which AD points appear and
calculate values of condensates at this point.  In the next section we study
what happen to the confinement of charges in the monopole point at non-zero
$\mu$ once we approach AD point.

First let us work out the AD values  of $m$, generalizing the
consideration~\cite{apsw}. Collision of two roots for $v$  means
that together with Eq.~(\ref{vaceq}) the derivative of its left-hand-side
should also vanish,
\begin{equation}
m-\frac{(-\det h)}{2} \, \frac{v}{\mu}- \frac{\Lambda_1^3}{4} \left(
\frac{\mu}{v}\right)^2=0, \qquad -(-\det h) + \Lambda_1^3 \left(
\frac{\mu}{v}\right)^3=0\;.
\label{vaceq1}
\end{equation}
This system is consistent only at three values of $m=m_{\rm AD}$,
\begin{equation}
m_{\rm AD}=\frac 3 4 \, \omega\, \Lambda_1\,(-\det  h)^{2/3},
\qquad \omega={\rm e}^{2\pi i n/3}\quad (n=0,\pm1)
\;,
\label{mad}
\end{equation}
related by $Z_3$ symmetry. The condensates at the AD vacuum are
\begin{eqnarray}
&&v_{\rm AD}=\omega \,\frac{\mu\,\Lambda_1}{(-\det
h)^{1/3}}\,,\nonumber\\[1mm]
&& u_{\rm AD}= \omega^{-1} \, \frac 3 4  \,  \Lambda_1^2\,  (-\det
h)^{1/3}\,,\nonumber\\[1mm]
&& s_{\rm AD}=\omega^{-1} \, \frac 1 4 \, \mu\Lambda_1^2 \,  (-\det h)^{1/3}\,.
\label{cad}
\end{eqnarray}

\section{Dyon condensates}\label{sec:mcd}

In this section we calculate various dyon condensates at three vacua of the
theory.
As it was discussed above holomorphicity allows us to find these condensates
starting from  consideration on the Coulomb branch in \ntwo near
the singularities associated with given massless dyon. Namely,
we calculate the monopole
condensate
near the monopole point, the charge condensate near the charge point and the
dyon
$(n_{m},n_{e})=(1,1)$ condensate near the point where this dyon is light.
Although
we start with small value of adjoint mass parameter $\mu$,  our results for
condensates are exact for any $\mu$.

\subsection{Monopole condensate.}

Let us start with calculation of the monopole condensate near the
monopole point. Near this point the effective low energy
description of our theory can be given in terms of \ntwo dual QED~\cite{sw}. It
includes light monopole hypermultiplet interacting with vector (dual) photon
multiplet in the same way as electric charges interact with ordinary photons.
Following Seiberg and Witten~\cite{sw} we write down the effective
superpotential
in the following form.
\beq
W= \sqrt{2}\,\tilde{M}MA_{D}+\mu\, U,
\label{mqed}
\eeq
where $A_{D}$ is a chiral neutral field (it is a part of \ntwo
dual photon multiplet in \ntwo theory) and $U\!\!=\!\Tr \Phi^2$. The
second  term
breaks \ntwo supersymmetry down to \none.

Variating  this superpotential with respect to $A_{D}$, $M$
and $\tilde{M}$ we find that $A_{D}=0$, i.e.  the monopole mass vanishes, and
\beq
\langle\tilde{M}M \rangle=-\frac{\mu}{\sqrt{2}}\frac{{\rm d}{u}\:}{{\rm
d}{a_{D}}}\,.
\label{mc}
\eeq
The condition $A_{D}=0$ means that the Coulomb branch near the
monopole point, where the monopole mass vanishes, shrinks to the single vacuum
state at the singularity while Eq.~(\ref{mc}) together with D flatness
condition (up to gauge transformation) $\tilde{M}=M$
determines the value of monopole condensate.

The non-zero value of monopole condensate ensures the
U(1) confinement for charges via the formation of
Abrikosov-Nielsen-Olesen  vortices. Let us work out
the r.h.s. of Eq.~(\ref{mc}) to determine the $\mu$
and $m$ dependence of the monopole condensate.
>From exact Seiberg-Witten solution \cite{sw} we have
\beq
\frac{{\rm d}{a_{D}}}{{\rm d}u}=
\frac{\sqrt{2}}{8\pi }\oint_\gamma \frac{{\rm d}x}{y(x)}\,.
\label{con}
\eeq
Here for $y(x)$  given by Eq.~(\ref{swu}) we use the form
\beq
y^2=(x-e_{0})(x-e_{-})(x-e_{+})\,.
\eeq
We get finally
\beq
\langle \tilde M M\rangle =2i\mu\left(u_{M}^2-\frac 3 4
m\Lambda_{1}^{3}\right)^{1/4}.
\label{mm}
\eeq

Now let us address the question: what happens with the monopole
condensate when we reduce $m$ and approach the AD point.
The AD point corresponds to particular
value of $m$ which ensures  colliding of monopole and charge
singularities in the $u$ plane. Near the monopole point we have
condensation of monopoles and confinement of charges while
near the charge point we have condensation of charges and
confinement of monopoles. As it was shown by 't~Hooft
these two phenomena cannot happen simultaneously~\cite{H}. The
question is:  what happen when monopole and charge points collide
in the $u$ plane?

The monopole condensate
 at the AD point is given by Eq.~(\ref{mm}) when
 $m_{AD}$ and $u_{AD}$ from Eqs.~(\ref{mad}) and (\ref{cad}) are substituted,
\beq
\langle \tilde M M\rangle_{AD}=0.
\eeq
We see that monopole condensate goes to zero at the AD point. Our derivation
above makes clear why it happens. At the AD point all three roots of $y^2$
become degenerate, $e_+=e_-=e_0$, so the monopole condensate which is
proportional to $\sqrt{e-e_0}$ naturally vanishes.

In the next subsection we
calculate the charge condensate in the charge point and show that it is also
goes to
zero as $m$ approaches its AD value~(\ref{mad}). Thus we interpret
the AD point as a deconfinement point for both monopoles
and charges.

\subsection{Charge and dyon condensates}

In this subsection we use the same method to calculate
values of charge and dyon condensate near charge and dyon points
respectively. We first consider $m$ above AD value (\ref{mad}) and
then continue our results to values of $m$ below $m_{AD}$. In
particular in the limit $m=0$ we recover $Z_{3}$ symmetry.

Let us start with the charge condensate. At $\mu=0$, $\det h=-1$ and large $m$
the
effective theory near the charge point
\begin{equation}
a=-\sqrt{2}\,m
\label{chp}
\end{equation}
on the Coulomb branch
is \ntwo QED.
The half of degrees
of freedom in  color doublets  becomes massless whereas the other
half  acquire large mass $2m$. These massless fields form one hypermultiplet
$\tilde Q_+, Q_+$ of charge particle in the effective electrodynamics.
 Once we add the mass term for the adjoint matter
 the effective superpotential near the charge point becomes
\beq
{\cal{W}}=\frac{1}{\sqrt{2}}\,\tilde{Q_{+}}Q_{+}A+m\,\tilde{Q_{+}}Q_{+}
 +\mu\, U
\eeq
Minimizing this superpotential we get condition (\ref{chp})
as well as
\beq
\langle\tilde{Q_{+}}Q_{+}\rangle= -\sqrt{2}\,\mu\,\frac{{\rm d} u}{{\rm d} a}\,.
\eeq
Now following the same steps which led us from (\ref{mc}) to
 (\ref{mm}) we get
\beq
\langle\tilde{Q_{+}}Q_{+}\rangle=2\,\mu\,(u_{C}^2-\frac 3 4
m\,\Lambda_{1}^{3})^{1/4}
\label{chc}
\eeq
Here $u_{C}$ is the position of charge point in the $u$ plane, $u_{C}=m^2$ at
large
$m$, see Eq.~(\ref{charge}).
Thus, at large $m$
\beq
\langle\tilde{Q_{+}}Q_{+}\rangle =2\,\mu \,m\,.
\label{chcl}
\eeq

Holomorphicity allows us to extend the result (\ref{chc}) to arbitrary $m$ and
$\det h$. So we can use Eq.~(\ref{chc}) to find  the charge condensate at
the AD
point. Using Eqs.~(\ref{mad}) and (\ref{cad}) we see that the charge condensates
vanishes in the AD point the same way the monopole one does. As it was
mentioned we interpret this as deconfinement for both charges and monopoles.

Similarly to the monopole condensate we can relate the
charge condensate with the quark one $v$,
\begin{equation}
  \label{ccV}
  \langle\tilde Q_+ Q_+ \rangle^2= v^2- \frac{\mu^3
      \Lambda_1^3}{v}
=v^2- 4\mu \, s\,,
\end{equation}
This expression differs from the one for the monopole condensate only
by sign.
The coincidence of the charge condensate with the quark one at large
$v$, i.e. at weak coupling is natural. The difference is due to
nonperturbative effects and  similar to the difference between
$a^2/2$ and  $u$ on
the Coulomb branch of the \ntwo theory. In strong coupling the
difference is not small, in particular, the charge condensate vanishes
in the AD point while the quark condensate remains finite.

Note that near the AD point we can consider an effective
superpotential which includes both
light monopole and charge fields simultaneously.
Such consideration leads to the same results for condensates.

Now let us work out the dyon condensate. More generally let us
introduce the dyon field $D_{i}\,$, $i=1,2,3$, which
stands for charge, monopole and $(1,1)$ dyon, $D_{i}=(Q_{+},M,D)$.
The arguments of the previous subsection
which led us to the result (\ref{mm}) for monopole condensate
gives for $\langle\tilde D_i \, D_i\rangle$
\beq
\langle\tilde D_i \, D_i
\rangle=2\,i\,\zeta_{i}\,\mu\left(u_{i}^2-\frac 3 4
\,m\,\Lambda_{1}^{3}\right)^{1/4},
\label{dyc}
\eeq
where $u_{i}$ is the position of the i-th point in the
$u$ plane and  $\zeta_{i}$ are  phase factors.

For the monopole condensate  at real values of $m$ larger than
$m_{AD}\!=\!(3/4)\Lambda_1(-\!\det h)^{2/3}$ Eq.~(\ref{mm}) gives
\beq
\zeta_{M}=1,
\eeq
while for charge from Eq.~(\ref{chc})
\beq
\zeta_{C}=-i.
\eeq
In fact one can fix the phase factor for charge imposing
the condition that the charge condensate should approach
the value $2m \mu$ in the large $m$ limit.
For dyon the phase factor is
\beq
\zeta_{D}= i\,.
\label{dph}
\eeq

At the AD point monopole and charge condensates go to zero, while
the dyon one remains non-zero, see (\ref{dyc}). Below the AD point
condensates are given by the same Eq.~(\ref{dyc}), but the phase
factors for charge and monopole can change its values~\footnote{Note
that quantum numbers of ``charge'' and ``monopole'' are also
changed, see \cite{BF}}. The dyon phase factor~(\ref{dph})
is not changing when we move through the AD point  because the dyon
condensate does not vanish at this point.

\section{The Argyres-Douglas point: how well the theory is defined}
\label{sec:walls}
As we discussed in Introduction in the AD point we encounter the
problem of not uniquely defined vacuum state. Indeed, when the mass
parameter $m$ approaches its AD value $m_{AD}$ we deal with two vacuum
states which can be distinguished by values of chiral condensates. It
is unlikely that the number of states with zero energy will  change
when we reach the AD point, it is very much similar to Witten index.
However, the continuity of chiral condensates we obtained above shows
that they are no longer parameters which differentiate two states once
we reach the AD point.

A natural possibility to consider is domain walls
interpolating between colliding vacua. In case of BPS domain walls
their tension is given by central charges,

\begin{equation}
T_{ab}=2\,|{\cal W}_{\rm eff}(v_a) - {\cal W_{\rm eff}}(v_b)|
\end{equation}
where $a$,$b$ label colliding vacua. The central charge here is
expressed via values of exact superpotential~(\ref{sup1}) in
corresponding vacua. The continuity of the condensate $v$ shows that
the domain wall becomes tensionless in the AD point. If such domain
wall were observable it could serve as a signal of two vacua.

Let us note one more interesting question. Namely the BPS tension
should obey the Picard-Fuchs equation providing the dependence
on the quark mass. The mass corresponding to the position
of the Argyres-Douglas point plays the role of the
"strong coupling singularity" like the monopole singularity
in the Seiberg-Witten solution. The generic structure
of the monodromies at the complex m plane should provide
the unique solution for the multiplet of tensions. It would
be very interesting to compare the solutions of Picard-Fuchs
equations with the tensions followed from the exact
superpotentials.

\section{Conclusions}\label{sec:disc}

We analyze monopole, charge and dyon condensates departing from
the Coulomb branch of the \ntwo theory. It results in the explicit relations
between
these condensates and those of the fundamental matter. The most
interesting
phenomenon occurs in the AD point: when the monopole and charge
vacua collide both the monopole and charge condensates vanish. We interpret
this
as a deconfinement of electric and magnetic charges in the AD
point.

Let us mention a relation to  finite-dimensional
integrable systems.
It was recognized that \ntwo theories are governed by  finite-dimensional
integrable systems. The integrable system responsible for \ntwo SQCD
was identified with the nonhomogenious XXX spin chain~\cite{gm}.
After perturbation to the \none theory the Hamiltonian of the integrable
system is expected to coincide with the superpotential of
corresponding \none theory. This has been confirmed by direct calculation
in the pure \ntwo gauge theory~\cite{vafa} as well in the  theory with
massive adjoint multiplet~\cite{dorey}. It would be very
interesting
to find a similar connection between spin chain Hamiltonians and
superpotentials in the \none SQCD. One more point to be clarified is a
meaning of the AD
point within approach based on integrability. Since the quark mass is
identified as a value of spin~\cite{gm} one could expect that at
particular values of spins corresponding to the AD mass XXX spin chain
has additional symmetries similar to superconformal ones.

I am  grateful to A. Vainshtein and A.Yung for the
collaboration on this issue.
The work  was partially supported by the grant INTAS-99-1705
and CRDF-RP1-2108.

\end{document}